\documentclass[aps, prd, twocolumn,print, showpacs, nofootinbib]{revtex4}
\usepackage{amsmath}
\usepackage{amsfonts}
\usepackage{amssymb}
\usepackage{txfonts}
\usepackage{mathrsfs}
\usepackage{graphicx}
\usepackage{epsfig}
\usepackage{dcolumn}
\usepackage{bm}
\newcommand*{\prdfigscale}{0.85}
\newlength{\prdcolwidth}
\newlength{\figwidth}
\newlength{\doublewide}
\begin{document}

\title{Constraining the ellipticity of millisecond pulsars with observed spin-down rates}
\author{Wen-Cong Chen$^{1,2\footnote{chenwc@pku.edu.cn}}$}

\affiliation{$^1$School of Science, Qingdao University of Technology, Qingdao 266525, China;
\\$^2$School of Physics and Electrical Information, Shangqiu Normal University, Shangqiu 476000, China
}

\date{18 May 2020}

\begin{abstract}
A spinning neutron star (NS) that is asymmetric with respect to its spin axis can emit continuous gravitational wave (GW) signals. The spin frequencies and their distribution of radio millisecond pulsars (MSPs) and accreting MSPs provide some evidences of GW radiation, and MSPs are ideal probes detecting high frequency GW signals. It is generally thought that MSPs originate from the recycled process, in which the NS accretes the material and angular momentum from the donor star. The accreted matter would be confined at the polar cap zone by an equatorial belt of compressed magnetic field fixed in the deep crust of the NS, and yields "magnetic mountain".  Based on an assumption that the spin-down rates of three transitional MSPs including PSR J1023+0038 are the combinational contribution of the accretion torque, the propeller torque, and the GW radiation torque, in this work we attempt to constrain the ellipticities of MSPs with observed spin-down rates. Assuming some canonical parameters of NSs, the ellipticities of three transitional MSPs and ten redbacks are estimated to be $\epsilon=(0.9-23.4)\times 10^{-9}$. The electrical resistivities of three transitional MSPs are also derived to be in the range $\eta=(1.2-15.3)\times 10^{-31}~\rm s$, which display an ideal power law relation with the accretion rate. The characteristic strains ($h_{\rm c}=(0.6-2.5)\times10^{-27}$) of GW signals emitting by these sources are obviously beyond the sensitivity scope of the aLIGO. We expect that the third-generation GW detectors like the Einstein Telescope can seize the GW signals from these sources in the future.
\end{abstract}

\maketitle

\section{INTRODUCTION} \label{S:intro}
The detections of gravitational wave (GW) mark the start of a new era of multimessenger astrophysics. So far, aLIGO had detected a number of GW events including the mergers of double black holes and double neutron stars (NSs) \citep{abbott2016,abbott2017}. Comparing with catastrophic mergers of compact objects, continuous high frequency GW signals would provide many valuable information on the evolution of the stars. Millisecond pulsars (MSPs) with an ellipticity should emit high frequency GW signals. The typical ellipticity that LIGO and VIRGO could detect continuous GWs would be $\epsilon <2\times 10^{-5}$ \citep{aasi2014,aasi2015,aasi2016}.

A spinning NS that is asymmetric with respect to its spin axis can also radiate continuous GWs signals. In principle, GW radiation would result in a spin-down of NSs. In observation, the lack of submillisecond pulsars may stem from the spin-down of GW radiation \citep{ander2005,bhat2017}. At present, the known fastest-spinning MSP PSR J1748-2446ad has a spin period of 1.396 ms, corresponding to a spin frequency of 716 Hz \citep{hess06}. For spin frequencies on the order of 700 Hz or
more \footnote{The magnetosphere with a minimum
magnetic field of $10^{8}$ G may also be responsible for the lack of
NSs with spin frequency larger than 700 Hz \citep{patr12}.}, the spin-down torque producing by the GW emission can be sufficiently strong to balance the accretion torque, resulting in a critical frequency like PSR J1748-2446ad \citep{chak03,chak08}. The 13 known accreting X-ray millisecond pulsars (AXMSPs) possess an average spin period of 3.3 ms, whereas that of recycled radio MSPs is 5.5 ms \citep{hess08}. The propeller torques during the Roche-lobe decoupling phase could interpret this apparent difference in spin period distributions between AXMSPs and radio MSPs \citep{taur12}. The spin frequencies of weakly magnetic ($\ll 10^{11}~\rm G$) accreting NSs are within a narrow range of 250 - 350 Hz. These spin similarities can easily explained by the GW radiation, which produces a spin-down rate with a strong spin frequency-dependence (see also the below Eq. \ref{gr}) \citep{bildsten1998}. The statistical analysis of the spin distributions shows that the accreting MSPs can be divided into two subpopulations, a slow population with a mean spin frequency of 300 Hz and a broad spread, and a fast one with an average spin
frequency of 575 Hz \citep{patr2017}. The spin frequencies of the fast population are within a very narrow range of frequencies (30 Hz), and the two subpopulations are separated at a frequency cut-point of 540 Hz. Various accretion torque models can not naturally account for the existence of a fast subpopulation. The GW radiation could play an important
role in producing the observed spin distributions of accreting MSPs, especially explaining the narrow
frequency range of the fast subpopulation and the frequency cut-point \citep{patr2017}. Therefore, the spin frequencies and their distribution of radio MSPs and accreting MSPs provide some evidences of GW radiation, and MSPs are ideal probes detecting high frequency GW signals. Apart from the spin frequencies and the distances, the characteristic strains of emitting GW depend on the ellipticity of NSs.  However, it is very difficult to constrain the ellipticity of NSs due to various uncertainties. For radio pulsars, the uncertainty of magnetic fields yield the uncertainty of magnetic dipole radiation. In accreting NSs, it is also impossible to untangle the contribution of GW radiation due to a high spin-up rate resulting from a high accretion rate. Transitional MSPs that undergo occasional transitions between radio pulsar and X-ray pulsar states provide an ideal opportunity to constrain the GW torque.

At present, three transitional MSPs including PSR J1023+0038 \citep{archibald2009,patr2014,stap2014}, XSS J12270$-$4859 \citep{de2013,bassa2014}, and IGR J18245$-$2452 \citep{papitto2013} were confirmed . In the radio pulsar state, three sources were observed to be spinning down \citep{archibald2013,papitto2013,ray2015}. Especially, timing of the radio pulsations in the high mode of the X-ray pulsar state of J1023 presented a precise measurement on the spin-down rate as $\dot{\nu}=-2.399\times 10^{-15}~\rm Hz\,s^{-1}$ \citep{archibald2013}. During the X-ray pulsar state, J1023 was detected the accretion powered pulsations \citep{archibald2015},  which was accompanied by a spin-down rate $\dot{\nu}=-3.041\times 10^{-15}~\rm Hz\,s^{-1}$ \citep{jaodand2016}. This spin-down rate is approximately 30\% higher than that in the radio pulsar state.

It was suggested that the increase in spin-down rate during the X-ray pulsar state originates from GW emission, which is due to the creation of a ¡°mountain¡± during the accretion \citep{haskell2017}. However, magnetic mountains relax resistively on a relatively long diffusive timescale $\sim 10^{8}~\rm yr$ after accretion ceases \citep{vigelius2009b}. J1023 should already experienced an accretion process before it evolve into radio MSPs, so it is still controversial whether the GW radiation can produce such a difference of spin-down rate between radio pulsar state and X-ray pulsar state. In the active state, J1023 shows a high state and a low state of X-ray, which were thought to be a rapid transition between the propeller phase and the radio pulsar phase \citep{camp2016}. Recently, a work argued that radio pulsar state and X-ray pulsar state correspond to the strong propeller with a low X-ray luminosity and the weak propeller with a high X-ray luminosity powered by accretion onto the NS, and the slightly increase of the magnetic torque causes an enhancement of spin-down rate \citep{ertan2018}.

In this work, an alternative model is proposed to interpret the difference of spin-down rate between radio pulsar state and X-ray pulsar state of J1023. The GW radiation torque would be always exerted on the NS in both states, while a strong propeller torque during the X-ray pulsar state results in an excess spin-down rate. Meanwhile, we attempt to constrain the ellipticities of MSPs with an observed X-ray luminosity and a spin-down rate. The paper is organized as follows. We describe different torques model
in Section 2. The model will be applied for three transitional MSPs and twelve redbacks with spin-down rates in Section 3. Finally, we make brief summary and discussion in Section 4.

\section{TORQUES MODEL}\label{Sec II}
In a low-mass X-ray binary, the NS would obtain the angular momentum from the accreted material,
and is spun up to a millisecond period. The accretion torque exerted on the NS is as follow
\begin{equation}
T_{\rm ac}=\dot{M}_{\rm acc}\sqrt{GMr_{\rm m}} \label{tacc},
\end{equation}
where $G$ is the gravitational constant, $\dot{M}_{\rm acc}$ is the accretion rate, $M$ is the NS mass.
In Eq. (\ref{tacc}), the magnetospheric radius $r_{\rm m}$ is
\begin{equation}
r_{\rm m}=1.1\times 10^{7} \dot{M}^{-2/7}_{13}M^{-1/7}_{1.4}\mu^{4/7}_{26}~\rm cm, \label{rm}
\end{equation}
where $\dot{M}_{13}=\dot{M}/10^{13}~\rm g\,s^{-1}$ is the mass inflow rate in the accretion disk, $M_{1.4}=M/1.4~\rm M_{\odot}$, $\mu_{26}=\mu/10^{26}~\rm G\,cm^{3}$ is the magnetic dipole moment of the NS. If the accretion efficiency of the NS is $\delta$, we have $\dot{M}_{\rm acc}=\delta\dot{M}$. Numerically, the spin-up rate yielding by the accretion torque can be written as
\begin{equation}
\dot{\nu}_{\rm ac}=7.2\times 10^{-17}\delta \dot{M}^{6/7}_{13}M^{3/7}_{1.4}I_{45}^{-1}\mu^{2/7}_{26}~\rm Hz\,s^{-1},
\end{equation}
where $I_{45}=I/10^{45}~\rm g\,cm^{2}$ is the moment
of inertia of the NS.

If the magnetospheric radius is greater than the corotation radius (at which the Keplerian angular velocity equals the spin angular velocity of the NS)
\begin{equation}
r_{\rm co} =\sqrt[3]{\frac{GMP^{2}}{4\pi^{2}}}=1.7\times 10^{6}M_{1.4}^{1/3}P_{-3}^{2/3}~\rm cm, \label{rco}
\end{equation}
the NS enters the so-called propeller phase, where $P_{-3}$ is the spin period of the NS in units of 1 ms. The propeller torque is given by \citep{ho2017}
\begin{equation}
T_{\rm pr}=-\frac{2\pi(1-\delta)\dot{M}r_{\rm m}^{2}}{P},
\end{equation}
which offers a spin-down rate as
\begin{equation}
\dot{\nu}_{\rm pr}=-1.2\times 10^{-15} (1-\delta)\dot{M}^{3/7}_{13}M^{-2/7}_{1.4}I_{45}^{-1}\mu^{8/7}_{26}P_{-3}^{-1}~\rm Hz\,s^{-1}. \label{nupr}
\end{equation}

If the magnetospheric radius is greater than the light cylinder radius
\begin{equation}
r_{\rm lc}=\frac{cP}{2\pi}=4.8\times 10^{6}P_{-3}~\rm cm,
\end{equation}
the NS will be visible as a radio pulsar, which could radiate strong radio emission by the magnetic dipole radiation. The torque providing by the magnetic dipole radiation is
\begin{equation}
T_{\rm md}=-\frac{16\pi^{3}\mu^{2}{\rm sin}^{2}\alpha}{3c^{3}P^{3}},
\end{equation}
where $\alpha$ is the inclination angle between the magnetic axis and the spin axis of the NS. Taking $\alpha=\pi/2$, the maximum spin-down rate by the magnetic dipole radiation can be written as
\begin{equation}
\dot{\nu}_{\rm md}=-9.7\times 10^{-15}I_{45}^{-1}\mu^{2}_{26}P_{-3}^{-3}~\rm Hz\,s^{-1}. \label{md}
\end{equation}

The magnetohydrodynamic (MHD) simulation \citep{payne2004} show that the accreted matter is confined at the polar cap zone by an
equatorial belt of compressed magnetic field fixed in the deep crust  \citep[see, e.g.][]{hameury1983,brown1998}. The corresponding
"magnetic mountain" gives rise to a quadrupole moment, and the ellipticity can be written as \citep{payne2004,melatos2005,vigelius2009a}
\begin{equation}
\frac{\epsilon_{\rm MHD}}{2\times 10^{-7}}=\frac{\bigtriangleup M}{M_{\rm c}}\left(1+\frac{\bigtriangleup M}{M_{\rm c}}\right)^{-1}, \label{mhd}
\end{equation}
where $M_{\rm c}\approx 2\times 10^{-5}~M_{\odot}$, $\bigtriangleup M$ is the accreted mass of the NS. Considering
$\bigtriangleup M\gg M_{\rm c}$, hence $\epsilon_{\rm MHD}\approx 2\times 10^{-7}$.

In principle, magnetic mountains relax resistively on a diffusive
timescale after accretion ceases \citep{vigelius2009b}. For an accreting MSP,
the equilibrium between the diffusion timescale and the accretion timescale leads to the establishment of a steady state,
in which the influx of accreted material equals the efflux of by the Ohmic diffusion. Therefore,
the saturation ellipticity of the accreting MSP is \citep{brown1998,vigelius2009a}
\begin{equation}
\epsilon={\rm min}\left\{\epsilon_{\rm MHD}, 5.1\times 10^{-9}\left(\frac{1.3\times 10^{-27}~\rm s}{\eta}\right)\frac{\dot{M}_{\rm acc}}{\dot{M}_{\rm Edd}}\right\}, \label{ellipticity}
\end{equation}
where $\eta$ is the electrical resistivity, $\dot{M}_{\rm Edd}=1.0\times 10^{18}M_{1.4}~\rm g\,s^{-1}$ is the Eddington accretion rate.

Considering the gravitational radiation of the NS with an ellipticity $\epsilon$, the torque receiving by the NS is
\begin{equation}
T_{\rm gr}=-\frac{1024\pi^{5}GI^{2}\epsilon^{2}}{5c^{5}P^{5}},
\end{equation}
where $c$ is the light velocity in vacuum. The spin-down rate producing by the gravitational radiation can be expressed as
\begin{equation}
\dot{\nu}_{\rm gr}=-2.7\times 10^{-14}I_{45}\epsilon_{-9}^{2}P_{-3}^{-5}~\rm Hz\,s^{-1}, \label{gr}
\end{equation}
where $\epsilon_{-9}=\epsilon/10^{-9}$.

If the NS is at the propeller phase, the total torque includes the accretion torque, the propeller torque, and the gravitational radiation torque. Therefore, its spin-down rate is
\begin{equation}
\dot{\nu}=\dot{\nu}_{\rm ac}+\dot{\nu}_{\rm pr}+\dot{\nu}_{\rm gr}. \label{nu1}
\end{equation}
The first term on the right hand side of Eq. (\ref{nu1}) is positive, while the other two terms are negative. However, the spin-down rate of the NS radiating radio emission is given by
\begin{equation}
\dot{\nu}=\dot{\nu}_{\rm md}+\dot{\nu}_{\rm gr}, \label{nu2}
\end{equation}
and both terms on the right hand side of Eq. (\ref{nu2}) are negative. Comparing Eqs. (\ref{md}) and (\ref{gr}), the gravitational radiation would dominate the spin evolution of the MSP with a spin period less than 1.7 ms for some typical parameters $I_{45}=\mu_{26}=\epsilon_{-9}=1$.

\begin{table}
\begin{center}
\centering \caption{ Some main observed parameters for three transitional MSPs.\label{tbl-1}}
\begin{tabular}{ccccc}
\hline\hline
Sources & $P $ & $\dot{\nu}_{\rm obs} $& $L_{\rm X} $&  References\\
        & (ms)& ($10^{-15}~\rm Hz\,s^{-1}$)&($10^{33}\rm erg\,s^{-1}$)& \\
\hline
PSR J1023+0038 & 1.688 & $-3.041$ & 3.0      & \cite{archibald2015,jaodand2016}\\
XSS J12270-4859 &1.686 & $-3.9$   & 4.2      &  \cite{roy2015,mira2020}        \\
IGR J18245-2452 & 3.932& $<0.013$ &1000     & \cite{papitto2013}  \\
\hline
\end{tabular}
\end{center}
\end{table}

\begin{figure}
\begin{center}
\includegraphics[width=3.6 in, height = 2.8 in,
clip]{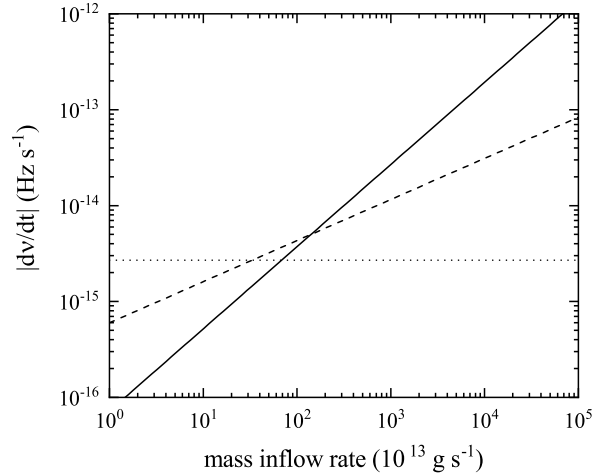} \caption{Relation between the spin-up rate (or spin-down rate) of MSPs and the mass inflow rate in the accretion disk. In this figure, we take $M_{1.4}=I_{45}=\mu_{26}=\epsilon_{-9}=1$, and $P_{-3}=2$. The solid, dashed, and dotted lines represent the the spin-up rate (or spin-down rate) of the accretion torque, the propeller torque, and the gravitational radiation torque, respecitvely.} \label{Fig1}
\end{center}
\end{figure}

Figure 1 illustrates the relation between the spin-up rate (or spin-down rate) producing by different torques and the mass inflow rate in the accretion disk. For an accreting MSP with spin period of 2 ms, the accretion torque dominate the spin evolution when $\dot{M}\geq 2\times 10^{15}~\rm g\,s^{-1}$. The propeller torque is dominant for a mass inflow rate in the range of $3\times 10^{14}-2\times 10^{15}~\rm g\,s^{-1}$. When the mass inflow rate declines to be lower than $3\times 10^{14}~\rm g\,s^{-1}$, the gravitational radiation torque of the NS with an ellipticity of $10^{-9}$ becomes the strongest one.

\begin{figure}
\begin{center}
\includegraphics[width=3.6 in, height = 2.8 in,
clip]{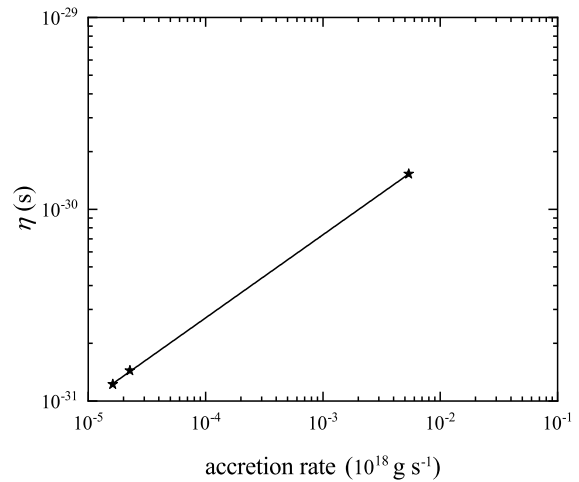} \caption{Electrical resistivity as a function of accretion rate (in units of Eddington accretion rate) for three transitional MSPs. The solid stars represent three sources, and the solid line denotes a power-law fit to the calculated results.} \label{Fig1}
\end{center}
\end{figure}

\section{APPLICATION FOR MSPS}\label{Sec IV}

\subsection{Transitional MSPs}

\begin{table*}
\begin{center}
\centering \caption{ Some derived parameters for three transitional MSPs.\label{tbl-2}}
\begin{tabular}{ccccccccccc}
\hline\hline
Sources & $\dot{M}_{\rm acc} $ & $\dot{M}_{13} $& $r_{\rm m}$ & $r_{\rm co}$ &$r_{\rm lc}$  &  $\dot{\nu}_{\rm ac}$& $\dot{\nu}_{\rm pr}$& $\dot{\nu}_{\rm gr}$& $\epsilon$& $\eta$\\
 & $(\rm 10^{13}~\rm g\,s^{-1})$ &   & ($10^{6}~\rm cm$) & ($10^{6}~\rm cm$)& ($10^{6}~\rm cm$) &$(10^{-16}~\rm Hz\,s^{-1})$ &$(10^{-15}~\rm Hz\,s^{-1})$& $(10^{-15}~\rm Hz\,s^{-1})$&$(10^{-9})$&$(10^{-31}~\rm s)$\\
\hline
J1023  & 1.62  & 10   &5.7  &  2.4  & 8.1 & 0.83  &$-1.6$  & $-1.52$ &0.9 &1.2\\
J12270 & 2.27  & 14   &5.2  &  2.4  & 8.1 & 1.1  &$-1.85$  & $-2.16$ &1.0  &1.4\\
J18245 & 540   & 540  &1.8  &  4.2  & 18.9& 158   &0        & $-15.79$&23.4 &15.3\\
\hline
\end{tabular}
\end{center}
\end{table*}

In this subsection, torques model are applied to three transitional MSPs. According to the X-ray luminosity, the accretion rate of the NS is given by
\begin{equation}
\dot{M}_{\rm acc}=\frac{L_{\rm X}R}{GM}=5.4\times10^{12}\left(\frac{L_{\rm X}}{10^{33}~\rm erg\,s^{-1}}\right)R_{6}M_{1.4}^{-1}~\rm g\,s^{-1},
\end{equation}
where $R=R_{6}10^{6}~\rm cm$ is the radius of the NS. If such an accretion rate equals the mass inflow rate in the disk, Eqs. (\ref{rm}) and (\ref{rco}) indicate $r_{\rm m}>r_{\rm co}$ for J1023 and J12270, i. e. these two sources should be in the propeller phase (J18245 is in the accretion phase, hence $\delta=1.0$). We now estimate the accretion efficiency in the propeller phase. Taking $I_{45}=\mu_{26}=1$, and $\alpha=\pi/4$ \citep{archibald2009}, $\dot{\nu}_{\rm md}\approx-1.0\times 10^{-15}~\rm Hz\,s^{-1}$ for J1023 during the radio pulsar state. In the X-ray pulsar state, the excess spin-down rate originates from the difference between the propeller torque and magnetic dipole torque (the accretion torque is ignored), so $\dot{\nu}_{\rm pr}\approx-1.64\times 10^{-15}~\rm Hz\,s^{-1}$. According to Eq. (\ref{nupr}), the accretion efficiency of J1023 can be estimate to be $\delta=0.16$. Normally, the NS only accretes a fraction $\delta=0.01-0.05 $ of the inflow mass in the accretion disk during the propeller phase \citep{cui1997,papitto2015,tsyg2016}. However, 3D MHD simulations indicated that
the accretion efficiencies during the propeller phase are in the range of 0.13 - 0.49 for a similar spin period \citep{lv2014}. Therefore, J1023 may provide an evidence of high accretion efficiency during the propeller phase. We then assume that J12270 also possesses a same accretion efficiency $\delta=0.16$. During the propeller phase, the mass inflow rate in the accretion disk is calculated by
\begin{equation}
\dot{M}=\frac{\dot{M}_{\rm acc}}{\delta}.
\end{equation}

Table I summarizes some main observed parameter for three transitional MSPs including the spin period, the frequency derivative, the X-ray luminosity. Taking $M_{1.4}=R_{6}=I_{45}=\mu_{26}=1$, we can obtain $\dot{\nu}_{\rm ac}$ and $\dot{\nu}_{\rm pr}$. Eq. (\ref{nu1}) yields $\dot{\nu}_{\rm gr}$, and then the ellipticity and electrical resistivity are derived from Eqs. (\ref{gr}) and (\ref{ellipticity}), respectively.  All derived parameters are presented in Table II. The ellipticities of three sources are estimated to be in the range $(0.9-23.4)\times 10^{-9}$, and the electrical resistivities are derived to be in the range of $(1.2-15.3)\times 10^{-31}~\rm s$. The observed data of J18245 originated from the duration of X-ray outburst. Such an anomalously high ellipticity is most likely related to high accretion rate during X-ray outburst. According to Eq. (\ref{rm}), a high mass inflow rate will result in a small magnetospheric radius. The X-ray spectrum features of J18245 including the broad emission line observed at an energy compatible with the Fe K$\alpha$ transition (6.4-6.97 keV) confirmed that it is an accretion-powered MSPs \citep{papitto2013}. Figure 2 plots the relation between the electrical resistivity and the accretion rate. Although the samples are rare, a relatively ideal power law fit emerges. The electrical resistivity $\eta=10^{-28.83\pm0.01}(\dot{M}_{\rm acc}/\dot{M}_{\rm Edd})^{0.433\pm0.003}$ s. When $\dot{M}_{\rm acc}/\dot{M}_{\rm Edd}=10^{-5}$, $\eta=10^{-31.00\pm0.03}~\rm s$, which is in good agreement with the minimum electrical resistivity $\eta_{\rm min}=10^{-30.5\pm5.0}~\rm s$ for transient accreting MSPs \citep{vigelius2010}.

\begin{table*}
\begin{center}
\centering \caption{Constraints on the ellipticity of twelve redbacks with observed spin-down rates \label{tbl-1}}
\begin{tabular}{ccccccccc}
\hline\hline
Sources & $\nu $ & $\dot{\nu} $& $d$  &$\dot{\nu}_{\rm md}$ & $\dot{\nu}_{\rm gr}$& $\epsilon_{-9}$& $h_{\rm c}$  &  References\\
        & (Hz)& ($10^{-15}~\rm Hz\,s^{-1}$)& (kpc) &($10^{-16}~\rm Hz\,s^{-1}$)&($10^{-15}~\rm Hz\,s^{-1}$)& &($10^{-27}$)& \\
\hline
PSR J1048+2339 &214.35 & $-1.38$ &2.0  & $-0.96$  &  $-1.28$ &10.25 & 1.0  &\cite{dene2016}\\
PSR J1227-4853 &592.99 & $-3.9$  &1.61 & $-20.23$  & $-1.88$ &0.97  & 0.9  &\cite{roy2015}\\
PSR J1431-4715 &497.03 & $-3.486$&1.56 & $-11.91$  & $-2.29$ &1.67  &1.1   &\cite{bate2015}\\
PSR J1622-0315 &260.05 & $-0.784$&1.14 & $-1.71$  & $-0.61$ &4.37   &1.1   &\cite{stra2019}\\
PSR J1723-2837 &538.87 & $-2.19$ &0.93 & $-15.18$  & $-0.67$ &0.74  &1.0   &\cite{crawford2013}\\
PSR J1740-5340A &273.95 & $-12.6$&2.2  & $-1.99$  & $-12.40$ &17.25 &2.5   &\cite{dami2001a,dami2001b}\\
PSR J1748-2021D&74.10 & $-3.22$  &8.24 & $-0.04$  & $-3.22$ &230.91 &0.6   & \cite{frei2008} \\
PSR J1816+4510 &313.17 &$-4.227$ &4.36 & $-2.98$  & $-3.93$ &6.95   &0.7   & \cite{stov2014}\\
PSR J1906+0055 &358.48 &$-0.427$ &4.48 & $-4.47$  & $-$     &$-$    & $-$  & \cite{stov2016}\\
PSR J1957+2516 &252.42 &$-1.748$ &2.66 & $-1.56$  & $-1.59$ &7.59   &0.8   & \cite{stov2016}\\
PSR J2215+5135 & 383.2 & $-4.9$  &2.77 & $-5.46$  & $-4.35$ &4.42   &1.0   & \cite{abdo2013}\\
PSR J2339-0533 & 346.71& $-1.695$&1.1  & $-4.04$  & $-1.29$ &3.09   &1.4   &  \cite{pletsch2015} \\
\hline
\end{tabular}
\end{center}
\end{table*}

Certainly, our estimation for the ellipticities of three transitional MSPs should have some
uncertainties, which arise from the magnetic dipole moment $\mu$ and the accretion efficiency $\delta$.
For J18245, according to $\dot{\nu}_{\rm obs}\ll\dot{\nu}_{\rm ac}$ and $\dot{\nu}_{\rm pr}=0$, we have $\dot{\nu}_{\rm gr}\approx\dot{\nu}_{\rm ac}$. Since the spin-up rate producing by the accretion is not sensitive to $\mu$ ($\dot{\nu}_{\rm ac}\propto\mu_{26}^{2/7}$) and $\delta=1$, the estimation for the ellipticity of J18245 is relatively reliable.
For J1023, $\dot{\nu}_{\rm obs}=-2.4\times 10^{-15}~\rm Hz\,s^{-1}$ in the radio phase, implying a maximum spin-down rate of magnetic dipole radiation $\dot{\nu}_{\rm md}=-2.4\times 10^{-15}~\rm Hz\,s^{-1}$, which can be used to derive a maximum magnetic dipole moment $\mu_{26}\approx1.55$ when $\alpha=\pi/4$. Such a magnetic dipole moment would
enhance $\dot{\nu}_{\rm pr}$ by a factor of 1.65. Therefore, $\dot{\nu}_{\rm gr}=0.48\times 10^{-15}~\rm Hz\,s^{-1}$ for J1023, which yields an ellipticity of $\epsilon=0.5\times 10^{-9}$. For J12270, according to Eq. (\ref{nupr}) $\dot{\nu}_{\rm pr}\propto\frac{1-\delta}{\delta^{3/7}}$, so $\dot{\nu}_{\rm pr}|_{\delta=0.05}=1.86\dot{\nu}_{\rm pr}|_{\delta=0.16}$.
When $\delta=0.05$, $\dot{\nu}_{\rm gr}=0.57\times 10^{-15}~\rm Hz\,s^{-1}$ for J12270, which also yields an ellipticity of $\epsilon=0.5\times 10^{-9}$. Therefore, the influence of the uncertainties of the magnetic dipole moment $\mu$ and the accretion efficiency $\delta$ on the ellipticity is not obvious because of the weak-dependence of the ellipticity for $\dot{\nu}_{\rm gr}$ ($\epsilon\propto\dot{\nu}_{\rm gr}^{1/2}$). Considering these uncertainties, the ideal power law relation between
the electrical resistivity and the accretion rate would slightly alter, while this change is not great due to a logarithmic coordinate.

\subsection{Redbacks}
Redbacks are a subpopulation of eclipsing MSPs with relatively more massive
companions ($\sim0.2-0.4~\rm M_{\odot}$) and orbital periods less than 1 day.
The regular radio eclipses imply a low-density, highly ionized
gas cloud enclosing the companions. These eclipsing material may arise from the companion winds evaporating by
the high-energy particles from MSPs \citep{heuv1988,rude1989}. At present, several models including
disrupted magnetic braking \citep{chen2013}, irradiation-induced cyclic mass transfer \citep{benv2014}, accretion-induced collapse \citep{smed2015}, and thermal and viscous instability in the accretion disks \citep{jia2015} were proposed to account for the formation of redbacks. Actually, some properties of transitional MSPs in the rotation-powered state are similar with redbacks. Once the mass inflow rates of these redbacks slightly increase, they will appear as candidates of transitional MSPs \citep{ertan2018}.

Although Roche-lobe overflow in redbacks may occur, the transferring matter is ejected by the
radiation pressure at the inner Lagrangian point during the radio-ejection phase of MSPs \citep{burd2001}.
Because of no mass accretion, hence redbacks with observed spin-down rates provide an opportunity to constrain the ellipticity.
Table III lists the observed the spin frequency, the spin frequency derivative, and the distance of twelve redbacks \footnote{Some data come from Australia Telescope National Facility Pulsar Catalog \citep{manc2005}.}.
Assuming that $I_{45}=\mu_{26}=1$, and magnetic inclination angle $\alpha=\pi/2$, we can obtain $\dot{\nu}_{\rm md}$ from Eq. (\ref{md}). Subsequently, Eq. (\ref{nu2}) yields $\dot{\nu}_{\rm gr}$. Finally, the ellipticity can be derived from Eq. (\ref{gr}). For PSR J1906+0055, its frequency derivative by the magnetic dipole radiation with $\mu_{26}=1$ and $\alpha=\pi/2$ exceeds the observed value. This result probably cause by an overestimation of magnetic field or magnetic inclination angle. The ellipticities of other ten sources are constrain to be $\epsilon=(0.74-17.25)\times 10^{-9}$. The calculated ellipticity ($\epsilon=2.3\times 10^{-7}$) of PSR J1748-2021D is obviously higher than other redbacks. This ellipticity is still in the reasonable scope ($\epsilon\approx2.0\times 10^{-7}$) of MHD simulation (see also Eq. \ref{mhd}). However, the magnetic field of PSR J1748-2021D with a spin period of 13.5 ms is most likely underestimated because it was not completely recycled. It is worth note that $\dot{\nu}_{\rm gr}$ of both PSR J1740-5340A and PSR J1748-2021D are 2-3 orders of magnitude higher than $\dot{\nu}_{\rm md}$.
Even if PSR J1748-2021D possess a relatively strong magnetic field $B=4\times10^{9}~\rm G$, $\mu_{26}=20$, so $\dot{\nu}_{\rm md}=-1.6\times10^{-15}~\rm Hz\,s^{-1}$. Comparing with the observed spin-down rate, it still requires an excess angular momentum loss mechanism such as GW radiation. Therefore, both PSR J1740-5340A and PSR J1748-2021D are important candidates detecting high frequency GW signals.

Our calculated ellipticities of redbacks also exist uncertainties, which originate from the uncertainties of the magnetic dipole moment of NSs. If $\mu_{26}=2$, the spin-down rate by the magnetic dipole radiation would increase by a factor of four due to $\dot{\nu}\propto \mu_{26}^{2}$. As a result, three sources including PSR J1227-4853, PSR J1431-4715, and PSR J1723-2837 would not require the GW radiation to account for the observed spin-down rate. If $\mu_{26}=3$, other three sources including PSR J1622-0315, PSR J2215+5135, and PSR J2339-0533 would also be ruled out the possibility of GW radiation. Adopting a relatively strong magnetic dipole moment $\mu_{26}=3$, the ellipticities of PSR J1048+2339, PSR J1740-5340A, PSR J1748-2021D, PSR J1816+4510, and PSR J1957+2516 can be estimated to be $\epsilon_{-9}=6.5, 16.1, 229.8, 4.4$, and 3.5, respectively. Comparing with Table III, the ellipticities of these five sources are not strongly affected by the magnetic dipole moment.

\subsection{Detectability of GW signals}

The characteristic strain of GW emitting by a NS can be written as \citep{abbott2007}
\begin{equation}
h_{\rm c}\approx 1.05\times 10^{-27}\epsilon_{-9}I_{45}\left(\frac{\nu_{\rm gw}}{1000~\rm Hz}\right)^{2}\left(\frac{1~\rm kpc}{d}\right),
\end{equation}
where $\nu_{\rm gw}=2/P$ is the GW frequency, $d$ is the distance of the source. The luminosity of GW radiation
\begin{equation}
L_{\rm gw}=\frac{2048\pi^{6}GI^{2}\epsilon^{2}}{5c^{5}P^{6}}\approx 1.1\times10^{36}\epsilon_{-9}^{2}I_{45}^{2}P_{-3}^{-6}~\rm erg\,s^{-1},
\end{equation}
so the timescale of GW radiation is
\begin{equation}
t_{\rm gw}\approx 5.7\times10^{8}\epsilon_{-9}^{-2}I_{45}^{-1}P_{-3}^{4}~\rm years.
\end{equation}
Therefore, for some typical parameters $\epsilon_{-9}=I_{45}=P_{-3}=1$, the detectability of GW signals emitting by the MSPs would sustain an enough long timescale.

Adopting the results of Table II, the characteristic strains of GW signals from J1023, J12770, and J18245 are $1.0, 0.8$, and $1.2\times 10^{-27}$ (we adopt a minimum distance of 1.8 kpc for J12770, see also \citep{de2014}). GW signals of eleven redbacks also show a similar tendency, the characteristic strains are in the range of $(0.6-2.5)\times10^{-27}$ (see also Table III). These signals are obviously lower than the strain sensitivity of the aLIGO that can detect the GW signals. However, they are not beyond the sensitivity scope of third-generation GW detectors like the Einstein Telescope. Assuming an observation time of 5 yr, the minimum ellipticity that is detectable by the Einstein Telescope at 90\% confidence level is about $\epsilon\approx10^{-9}$ for a GW frequency of $\nu_{\rm gw}\sim 1000~\rm Hz$ \citep{magg2020}. Even if we adopt a relatively strong magnetic dipole moment $\mu_{26}=3$, the characteristic strains of GW signals from PSR J1048+2339, PSR J1740-5340A, PSR J1748-2021D, PSR J1816+4510, and PSR J1957+2516 are in the range of $(0.4-2.3)\times 10^{-27}$, which are still in the strain sensitivity of the third-generation GW detectors like the Einstein Telescope.

\section{CONCLUSION AND DISCUSSIONS}\label{Sec V}

Three transitional MSPs and twelve redbacks were reported to be spinning down, and the spin-down rate of J1023 during X-ray pulsar state is faster than that in radio pulsar state. In this work, we propose that the "magnetic mountain" induced by the accretion can cause the GW radiation, and the excess spin-down rate of J1023 during the accretion originates from the difference between the propeller torque and the magnetic dipole radiation torque. To account for the observation of J1023, the accretion efficiency $\delta=0.16$ in the propeller phase. Assuming that two transitional MSPs possess a same accretion efficiency in the propeller phase (J18245 is in the accretion phase), and taking $M_{1.4}=R_{6}=I_{45}=\mu_{26}=1$, the ellipticities of three sources are estimated to be $\epsilon=(0.9-23.4)\times 10^{-9}$. Meanwhile, the ellipticities of ten sources among twelve redbacks with observed spin-down rates are also constrained to be $\epsilon\approx(0.7-17.3)\times 10^{-9}$. Our constraints are in good agreement with the minimum ellipticity of $10^{-9}$ for MSPs given by \cite{woan2018}. These ellipticities are also nice within the scope constraining by possible equations of state of NS \citep{owen2005,john2013}.

Based on the saturation ellipticity given by \cite{vigelius2009a}, the electrical resistivities of three transitional MSPs are derived to be $\eta=(1.2-15.3)\times 10^{-31}~\rm s$. There exist a nicely power law relation between the electrical resistivity and the accretion rate as $\eta=10^{-28.83\pm0.01}(\dot{M}_{\rm acc}/\dot{M}_{\rm Edd})^{0.433\pm0.003}$ s. This power law relation is consistent with the minimum electrical resistivity ($\eta=10^{-28\pm 4}(\dot{M}_{\rm acc}/\dot{M}_{\rm Edd})^{0.5\pm0.2}$ s) for transient accreting MSPs \citep{vigelius2010}.

Although the torques model is successful in explaining the difference of spin-down rates between X-ray pulsar state and radio pulsar state of J1023. However, the torques model strongly depend on the magnetic dipole moment, and the accretion efficiency, hence our constraints on the ellipticity of MSPs contain some uncertainties. In three transitional MSPs, the influence of magnetic dipole moment for J18245 can be neglected, while an inferred maximum $\mu$ would yield an ellipticity of $0.5\times 10^{-9}$ for J1023, and a relatively low accretion efficiency $\delta=0.05$ also produce an ellipticity of $0.5\times 10^{-9}$ for J12270. Furthermore, the ellipticities of five sources among twelve redbacks decrease by a maximum factor of 2 even if a strong magnetic dipole moment $\mu_{26}=3$ is adopted. Therefore, our estimations for the ellipticities remain marginal reliability. As a result, there exist a possibility that these sources can be detected by the third-generation GW detectors like the Einstein Telescope. In particular, two redbacks PSR J1740-5340A and PSR J1748-2021D are important candidates detecting high frequency GW signals.

There exist three promising observational checks whether the additional spin-down rate during the X-ray pulsar state of J1023 arises from the propeller torque. First, the excess spin-down rate should sharply vanish when the accreting MSP move to radio pulsar state. On the contrary, it would disappear on a specific timescale if it results from the GW radiation \citep{haskell2017}. Second, the characteristic strains of GW signals emitting in the X-ray pulsar and the radio pulsar states should have an approximately same strength. Third, the measured braking index during the radio pulsar state should be $3<n<5$ because the braking torques are combination between magnetic dipole radiation and GW radiation \citep{chen2016}, like PSR J1640-4631 \citep{archibald2016}.

The detection of GW for the accreting MSPs is very significant. The angular momentum loss rate by the GW radiation can be derived according to the measurement of GW amplitude and frequency, and then the accretion torque of the disk can be also inferred \citep{watt2008}. Therefore, the detection of GW would provide an important constraint on the accretion disk model and the magnetic field of the MSP. However, the GW signals emitting by MSPs with an ellipticity of $10^{-9}$ can not be detected by the aLIGO.
We expect that the third-generation GW detectors like the Einstein Telescope can seize the GW signals of some accreting MSPs in the future.

\acknowledgements We thank the referee for a very careful reading and constructive comments that have led to the improvement of the manuscript. This work was partly supported by the National Natural Science Foundation
of China (under grant number 11573016, 11733009), Program for Innovative Research Team (in Science and Technology)
in University of Henan Province, and China Scholarship Council.

\end{document}